\documentstyle[aps,preprint]{revtex}
\begin{document}
\draft
\title{Description of superdeformed nuclei in the interacting boson model}
\author{S. Kuyucak$^1$, M. Honma$^2$ and T. Otsuka$^3$}
\address{$^1$Department of Theoretical Physics,
Research School of Physical Sciences,\\
Australian National University, Canberra, ACT 0200, Australia\\
$^2$Center for Mathematical Science, University of Aizu, Tsuruga,
Ikki-machi, Aizu-Wakamatsu, Fukushima 965, Japan\\
$^3$Department of Physics, University of Tokyo, Bunkyo-ku, Tokyo 113, Japan}
\date{\today}
\maketitle

\begin{abstract}
The interacting boson model is extended to describe the spectroscopy of
superdeformed bands. Microscopic structure of the model in the second minimum
is discussed and superdeformed bosons are introduced as the new building
blocks.
Solutions of a quadrupole Hamiltonian are implemented through the $1/N$
expansion method. Effects of the quadrupole parameters on dynamic moment
of inertia and electric quadrupole transition rates are discussed and the
results are used in a description of superdeformed bands in the Hg-Pb and Gd-Dy
regions.
\end{abstract}

\section{INTRODUCTION}

Spectroscopy in the superdeformed minimum has reached a certain level of
maturity to justify a phenomenological analysis of the available data (see
\cite{jan91,nol94} for recent reviews).  Such an approach would be useful in
systematizing the data and would also provide a complimentary perspective
to the more microscopic theories. For this purpose, we use the interacting
boson model (IBM) \cite{iac87} which has been established as one of the
simplest and most versatile collective models. It has been especially
successful in correlating spectroscopic data in deformed nuclei in terms of a
few parameters of a quadrupole Hamiltonian \cite{cas88}.

Microscopic study of the nucleon pair structure in the superdeformed well
\cite{hon92} indicates that, compared to the deformed nuclei, they have
about three times more active pairs of nucleons (bosons), and the $L=4$ pair
($g$ boson) plays a much more significant role. As numerical diagonalization of
an $sdg$-IBM Hamiltonian is not possible for more than 10 bosons, one needs
alternative methods of solution to apply the IBM to superdeformed nuclei.
Here, we use the angular momentum projected mean field
theory which leads to a $1/N$ expansion for all matrix elements of
interest \cite{kuy88}. Accurate representation of high-spin states in the
$1/N$ expansion formalism requires terms to order $(L/N)^6$ which have been
obtained recently using computer algebra \cite{kuy95}. The extended formalism
provides an analytical method for analysis of superdeformed states which is
both accurate and efficient.

The plan of the paper is as follows. After reviewing the microscopic basis
of the IBM for superdeformed nuclei, we introduce the $1/N$ expansion formalism
and
discuss its recent extension to higher orders. We then use the $1/N$ expansion
formulas for a quadrupole Hamiltonian to study systematic features of
dynamic moment of inertia and $B(E2)$ values. The results are used in a
description the superdeformed bands in the Hg-Pb  and Gd-Dy regions.

\section{MICROSCOPIC BASIS}

In this section, we study the typical structure of  strongly deformed states
and investigate the relation between the superdeformation and the IBM
 based on \cite{hon92}.
For this purpose, we use the Nilsson + BCS model  with particle number
projection. Superdeformed states can be characterized as ground states  in a
superdeformed potential well which is separated  from the normal one by a
potential barrier. For such ground-like states which show strongly collective
nature, this model seems to work well. Using the experimental deformation
parameters and electric transition probabilities (or moments) as input, one can
obtain reasonable wave functions. These wave functions are analyzed from the
viewpoint of  collective nucleon pairs,  which leads to a natural extension of
the usual IBM.

We briefly summarize the formulation of the  Nilsson +
particle-number-conserving BCS model. The single particle orbits in a deformed
potential are described well by the Nilsson Hamiltonian \cite{nil55}
\begin{eqnarray}
H_{\rm Nilsson} &=& -\frac{\hbar^{2}}{2m} \nabla ^2
              +\frac{m}{2}\omega_{0}^{2}r^{2}
                 [1-\frac{4}{3}\delta P_{2}(\cos\theta)] \nonumber \\
       &-& 2\hbar \omega_{0} \kappa \mbox{\boldmath $l$} \cdot
        \mbox{\boldmath $s$} -\hbar \omega_{0} \kappa \mu
      (\mbox{\boldmath $l$}^{2}-\langle \mbox{\boldmath $l$}^{2} \rangle_{N}),
\end{eqnarray}
where $\delta$ is the deformation parameter and  $P_{2}(\cos\theta)$ denotes
the Legendre polynomial.
The term $\langle \mbox{\boldmath $l$}^{2} \rangle_{N} = \frac{1}{2}N(N+3)$ is
the expectation value of $\mbox{\boldmath $l$}^{2}$
averaged over one major shell with the principal quantum number $N= 2n+l$.
The value of the oscillator frequency for a mass-A nucleus is determined
from $\hbar\omega_{0} = 41A^{-\frac{1}{3}}$, which is about 7.1 MeV for
the superdeformed nuclei in the Hg-Pb region ($A \sim$ 190) and
7.7MeV for those in the Dy-Gd region ($A \sim$150).
We use the usual values for the parameters $\kappa$ and $\mu$
which are 0.0637, 0.60 for proton orbits and 0.0637, 0.42 for neutron orbits,
respectively \cite{gus67}.
In order to include short range correlations, the monopole pairing
interaction is added to the Nilsson hamiltonian \cite{nil61}
\begin{equation}
H = H_{\rm Nilsson} + G P^{\dagger} P,
\end{equation}
where $G$ denotes the pairing strength parameter, and
\begin{equation}
P^{\dagger} = \sum_{k>0} c_{k}^{\dagger} c_{\bar k}^{\dagger} \ ,
\end{equation}
is a pair creation operator.
Here $c_{k}^{\dagger}$ stands for the creation operator of a nucleon
in the spherical single particle orbit $k$,
and $\bar k$ denotes the time reversed state of $k$.
This Hamiltonian is solved by the variation using a BCS wave function
\begin{equation}
\mid \Psi \rangle = \prod_{\alpha>0} (u_{\alpha} + v_{\alpha}
  a_{\alpha}^{\dagger} a_{\bar \alpha}^{\dagger}) \mid 0 \rangle,
\end{equation}
where $a_{\alpha}^{\dagger}$ denotes the creation operator
for a nucleon in the deformed canonical (Nilsson) orbit labeled by $\alpha$.
The particle number conservation has been found to be important in the case of
weak pairing correlations and also for moments of inertia of high spin states
in the cranking calculation of the superdeformed states \cite{dud88,shi90}.
Thus we carry out the particle number projection before variation
according to the method given in Ref. \cite{egi82}.
The solution corresponds to the minimum of the number projected energy
\begin{equation}
E^{P}[\Psi] = \frac{\langle \Psi \mid H P^{N} \mid \Psi \rangle}
                   {\langle \Psi \mid P^{N} \mid \Psi \rangle},
\end{equation}
where $P^{N}$ denotes the particle number projection operator.

The deformation parameters of the superdeformed states in the Hg-Pb (Dy-Gd)
region are given by $\delta \sim$ 0.40 (0.50), which is equivalent to the
axis ratio of 5:3 (2:1). Because of this strong deformation, it is insufficient
to take only one active major shell and take into account
 the corrections due to the
core-polarization effect through renormalization in one major shell.
 Thus we first seek a suitable model space for
description  of superdeformed states. For simplicity, we turn off the
pairing force which is not important for this purpose. We take $^{194}$Hg
($N$=114, $Z$=80) and $^{152}$Dy ($N$=86, $Z$=66) as  examples of the Hg-Pb
and the Dy-Gd regions, respectively.

In order to define the model space  necessary for description of
superdeformed states,  we utilize the intrinsic quadrupole moment $Q_0$.
The intrinsic quadrupole moment is calculated in the space of all spherical
orbits up to the principal  quantum number $N=N_{\rm max}$. Then $N_{\rm max}$
is increased until the value of $Q_0$ is saturated to a good extent.
{}From this procedure we obtain $N_{\rm max}=12$. The corresponding values of
$Q_0$ for proton and neutron orbits are 19{\it b} and 29{\it b} for $^{194}$Hg,
and 19{\it b} and 27{\it b} for $^{152}$Dy, respectively.
Note that the experimental values of $Q_0$ are $18\pm 3~eb$ for $^{152}$Dy
\cite{ben91} and  $17\pm 2~eb$ for $^{194}$Hg \cite{hug94}, which are
consistent  with the present results if we take the bare charges, $e_{\rm p}=1$
and $e_{\rm n}=0$.

We now consider the inert core of superdeformed states. The Nilsson wave
function is obtained by putting all nucleons in the Nilsson  orbits from the
bottom. One Nilsson orbit can be expanded as a linear combination  of many
spherical harmonic oscillator orbits,  and the square of expansion coefficients
gives the occupation probability of each spherical orbit. We expand all the
occupied Nilsson orbits and sum up all the occupation  probabilities which
belong to the same spherical orbits, to obtain the total occupation
probability for a given spherical harmonic oscillator basis. Due to the strong
quadrupole field, one Nilsson orbit  spreads over many spherical orbits. Thus
the orbits with very high single particle energy can  gain some finite
occupation probabilities,  while the occupation of the orbits with small single
particle energy may become incomplete. Nevertheless several lower spherical
orbits are occupied almost completely and can be considered as  a new inert
spherical core for the superdeformed states. Note that we do not take the usual
``hole'' picture as it is meaningful only for states whose configuration are
well described within one major shell.

First consider the case of $^{194}$Hg. In Fig.~\ref{fig1}  the occupation
probability of each spherical  harmonic oscillator orbit is shown  for
neutrons (a) and protons (b). The orbits are ordered according to their single
particle  energy at $\delta=0$ as $1s_{1/2}$, $1p_{3/2}$, $\cdots$. The case of
$\delta=-0.13$ which simulates the deformation of normal  oblate states is also
shown for comparison.
For normal deformation, it is seen from Fig.~\ref{fig1}-a that the
 occupation of
the proton orbits is  almost complete at $2d_{5/2}$ ($Z=64$), while the
occupation probability  is almost vanished for orbits above $Z=82$.
These results suggest that we can consider the $Z=64$ subshell
as an inert core and three valence orbits ($2d_{3/2}$, $3s_{1/2}$ and
$1h_{11/2}$) as active.
This gives the valence proton number as $Z_{v}=16$.
In the case of $\delta=0.40$, the proton orbits are almost completely
occupied up to $Z=50$. Above $Z=50$, the occupation probability
drops suddenly though it remains about 10$\%$ over many orbits.
Clearly, one should incorporate the contributions of these high
energy orbits. Thus it is reasonable to take the $Z=50$ spherical inert core
and  include quite many orbits above there as active valence orbits. In this
case the valence proton number becomes $Z_{v}=30$.

In the same way, it can be
seen form Fig.~\ref{fig1}-b that  the spherical inert core for neutron orbits
are  $N=100$ ($N_{v}=14$) and 82 ($N_{v}=32$)  for $\delta=-0.13$ and
0.40, respectively. The active valence orbits for $\delta=-0.13$ are
$2f_{5/2}$, $3p_{3/2}$,  $3p_{1/2}$ and $1i_{13/2}$,  while for $\delta=0.40$
it is still insufficient to include  only two or three major shells.

We can see a similar behaviour of occupation probabilities  in the
wavefunctions of $^{152}$Dy,  which is shown in Figs. \ref{fig1}-c for
proton  and \ref{fig1}-d for neutron orbits. In these figures two cases of
$\delta=0.50$ (superdeformed state)  and 0.25 (normal prolate state) are
compared. It is clear that $Z=50$ and $N=82$ inert cores are good for  normal
states, while $Z=40$ and $N=50$ cores are suitable for  superdeformed states.
Thus the inert core of superdeformed states becomes much smaller  than that of
normal states in both the Hg-Pb and Dy-Gd regions.

{}From the viewpoint of the IBM, the number of bosons is determined by  half of
the number of valence nucleons. Because of the small inert core the number of
bosons increases  significantly for superdeformed states  in comparison with
that in the usual IBM. In fact in the case of $^{194}$Hg, the boson number
in the usual IBM is  $N_{\rm normal}=(82-80)/2+(128-114)/2=8$ by taking
the usual hole picture,  and $N_{\rm normal}=(80-64)/2+(114-100)/2=15$ with the
particle picture  mentioned above. On the other hand, the number of bosons for
superdeformation becomes $N_{\rm super}=(80-50)/2+(114-82)/2=31$.
Similarly, in the case of $^{152}$Dy,  $N_{\rm normal} = (66-50)/2 +
(86-82)/2 = 10$  while $N_{\rm super} = (66-40)/2 + (86-50)/2 = 31$. In
general, $N_{\rm super}$ is about three times larger  than $N_{\rm normal}$.
Note that the number of proton bosons and neutron bosons are  close in these
two cases, and this approximate equality seems to  be a general tendency of the
superdeformed states. This result can be naturally understood  since equal
numbers of valence protons and  neutrons maximizes the attractive
proton-neutron interaction.

Next we consider the effects of pairing correlations  on the structure of
wave functions of superdeformed states. The strength parameter $G$ of the
pairing interaction  should be chosen depending on the model space.
Since the value of $G$ for such a large space is not known empirically,
we first describe normal states  within the extended valence space,
and determine the value of $G$ by requiring that  the pairing gap ${\it
\Delta}$ takes a reasonable value. For $^{194}$Hg the value of $G$ has turned
out to be 0.06MeV  which gives ${\it \Delta}\sim 1$ MeV. Using this value, we
investigate the effect of the pairing correlations  on the structure of valence
wave functions. For this value of $G$, the gap for superdeformed states becomes
about $\Delta =0.5$ MeV for both proton and neutron orbits.
In contrast to normal deformed states, which are sensitive to changes in values
of $G$, the superdeformed states are almost insensitive to $G$ values
(the intrinsic quadrupole moment  and the occupation probabilities
change very little). Thus the following discussion about the structure of
valence wave function of superdeformation
 is almost independent of  pairing correlations.

We can investigate the relation between the superdeformation  and the IBM by
analyzing valence wave functions  from the viewpoint of collective nucleon
pairs. Since the bosons in the IBM are understood as images of these
pairs,  such an analysis is essential in establishing a microscopic basis for
the super IBM. We consider $^{194}$Hg as an example.

The Nilsson + particle-number-conserving BCS wave function can be expressed
as the condensed state of coherent Cooper-pairs in the deformed potential
\cite{ots86}
\begin{equation}
 P^{N} \mid \Psi \rangle \propto ({\it \Lambda}_{\pi}^{\dagger})^{N_{\pi}}
                   ({\it \Lambda}_{\nu}^{\dagger})^{N_{\nu}}
                    |0 \rangle ,
\end{equation}
acting on the inert core $|0 \rangle$. In this expression,
${\it \Lambda}_{\pi}^{\dagger}$ (${\it \Lambda}_{\nu}^{\dagger}$)
denotes the creation operator of a Cooper-pair in proton (neutron) orbits and
$N_{\pi}$ ($N_{\nu}$) means half of the valence proton (neutron) number.
These ${\it \Lambda}$-pairs can be decomposed into a linear combination of
collective nucleon pairs with good angular momenta
\begin{equation}
 {\it \Lambda}^{\dagger} = x_{0}S^{\dagger}
                   + x_{2}D_{0}^{\dagger}
                   + x_{4}G_{0}^{\dagger}+\cdots ,
\end{equation}
where $S^{\dagger}$, $D^{\dagger}$, $G^{\dagger}$, $\cdots$ denote  the
collective nucleon pairs with  spin-parity $J^{\pi}=$ $0^{+}$, $2^{+}$,
$4^{+}$, $\cdots$ and the $x_{J}$'s  are amplitudes. The probability of each
pair  in the ${\it \Lambda}$-pair is given by the square of each amplitude,
and is listed in Table~\ref{table1} for  two cases of $\delta=-0.13$ and
$\delta=0.40$. It is well known that in the case of normal deformation  the
dominant components are the $S$- and $D$-pairs  \cite{ots82,bes82}.
In fact, these two components account for 100$\%$ probability
in the case of $\delta=-0.13$. In the case of $\delta=0.40$,  the total
probability of  the $S$- and $D$-pairs is about 80$\%$ and we can
conclude that  these pairs are still dominant in the ${\it \Lambda}$-pair.
However the probability of the $G$-pair is now sizable,  and it can no
longer be neglected in a detailed description of high-spin states.
It should be noted that the ratio of the $S$-pair to the other pairs
is quite similar to that of $s$-boson to the other bosons in the  SU(3) limit
of the IBM, which are shown in the same table. This suggests that the SU(3)
limit of the $sdg$-IBM could provide a reasonable phenomenological framework
for superdeformed states.

To summarize the microscopic results,  we emphasize two important points
 for the
description of superdeformed bands in the IBM:  One is the significant increase
in the boson number,  and the other is the importance of g-bosons. In addition,
it has been found that  the bosons for superdeformed states carry the
collectivity over many major-shells and that  the SU(3) limit is a reasonable
starting point.

\section{1/N EXPANSION FOR SUPER IBM}

A simultaneous description of the spectroscopy of normal and superdeformed
states requires rather complicated wave functions, therefore we focus on the
latter here and leave the complete picture for future work. We introduce
the superbosons ${\bf s}, {\bf d}, {\bf g}$ as the boson images of the $S, D,
G$ collective nucleon pairs in the superdeformed well (bold face notation is
used for super bosons to distinguish them from the normal ones). The quadrupole
Hamiltonian for this system of bosons has the form
\begin{equation}
H=-\kappa {\bf Q} \cdot {\bf Q},
\label{ham}
\end{equation}
where the quadrupole operator is defined as
\begin{equation}
{\bf Q} =
[{\bf s}^\dagger {\tilde {\bf d}} + {\bf d}^\dagger {\tilde {\bf s}} ]^{(2)} +
q_{22} [{\bf d}^\dagger {\tilde {\bf d}} ]^{(2)} +
q_{24} [{\bf d}^\dagger {\tilde {\bf g}} +
{\bf g}^\dagger {\tilde {\bf d}} ]^{(2)} +
q_{44} [{\bf g}^\dagger {\tilde {\bf g}} ]^{(2)}.
\label{qsdg}
\end{equation}
Here brackets denote tensor coupling of the boson operators and
$\tilde b_{lm}=(-1)^{m}b_{l-m}$. The parameters $q_{jl}$ in Eq. (\ref{qsdg})
determine strengths of boson interactions relative to the $s-d$ coupling.
Since the SU(3) limit is used as a reference
point in the rest of the paper, we quote the values for the quadrupole
parameters in this limit; $q_{22}=-1.242$, $q_{24}=1.286$, $q_{44}=-1.589$.
As stressed in the introduction, numerical diagonalization of this Hamiltonian
for $N \sim 30$ bosons is not possible even on a supercomputer.
The large number of bosons are, however, an advantage for the analytic
$1/N$ expansion technique which we employ here for solving the Hamiltonian
Eq. (\ref{ham}).
The $1/N$ expansion method has previously been discussed in detail \cite{kuy88}
and the recent extensions to higher orders are given in Ref.~\cite{kuy95}.
Therefore, we give only a short account of the formalism here, focusing
mainly on the accuracy of the results for high-spin states.
The starting point of the $1/N$ calculations is the boson condensate
\begin{equation}
|N,{\bf x}\rangle =(N!)^{-1/2}({\bf b}^\dagger)^N|0\rangle,\quad
{\bf b}^\dagger= x_0 {\bf s}^\dagger + x_2 {\bf d}_0^\dagger +
x_4 {\bf g}_0^\dagger,
\end{equation}
where $x_l$ are the mean field amplitudes to be determined by variation after
projection (VAP) from the energy expression
\begin{equation}
E_L=\langle N,{\bf x}| H P^L_{00} | N,{\bf x}\rangle /
\langle N,{\bf x}| P^L_{00} | N,{\bf x}\rangle.
\end{equation}
Here $P^L_{00}$ denotes the projection operator.
The resulting energy expression is a double expansion in $1/N$ and
$\bar L=L(L+1)$, and has the generic form
\begin{equation}
E_L = N^2 \sum_{n,m} {e_{nm}\over (aN)^m}
\Bigl({\bar L \over a^2N^2}\Bigr)^n,
\label{me1}
\end{equation}
where $a=\sum_l \bar l x_l^2$ and the expansion coefficients $e_{nm}$ involve
various quadratic forms of the mean fields $x_{l}$, e.g.,
$e_{00}=(\sum_{jl} \langle j0 l0|20 \rangle q_{jl} x_j x_l)^2$.
The coefficients $e_{nm}$ have recently been derived
up to the third order, $(\bar L/N^2)^3$, using computer algebra \cite{kuy95}.

Another observable of interest in the study of superdeformed states is the
$E2$ transitions. Assuming that the quadrupole transition operator is the same
as in the Hamiltonian, i.e. $T(E2)=e{\bf Q}$ where $e$ is an effective boson
charge, the $E2$ matrix elements are given by
\begin{equation}
\langle L \parallel T(E2) \parallel L-2 \rangle =
e N \hat L \langle L0\, 20|L-2\ 0\rangle \bigl[m_1 + m_2 L(L-1)/N^2\bigr]
\label{e2}
\end{equation}
where $\hat L = [2L+1]^{1/2}$ and the coefficients $m_n$ are given in
Ref.~\cite{kuy95}. The first term in Eq. (\ref{e2}) gives the familiar
rigid-rotor result. The second term is negative and is responsible for the
boson cutoff effect in $E2$ transitions.

Before applying the 1/$N$ expansion results, we compare them with those
obtained from an exact diagonalization of the Hamiltonian.
Diagonalization is carried out for $N=10$ which is the maximum possible boson
number for this purpose.
The quadrupole parameters $q_{22}, q_{24}, q_{44}$ are scaled down from their
SU(3) values with $q=0.7$ which gives an adequate parametrization for the Hg-Pb
region.
Fig.~\ref{fig2}-a compares exact results for the dynamic moment of inertia
${\cal J}^{(2)}$ (circles) with three different $1/N$ calculations.
The solid line shows the third order VAP results which is seen to follow the
exact results very accurately. The second order VAP (dotted line) and
the third order VBP results (dashed line) break down around spin $L\sim 2N$.
Hence for description of high-spin states, the third order $1/N$ expansion
formulas with VAP seem to be both necessary and sufficient.
In Fig.~\ref{fig2}-b, the exact $E2$ transition matrix elements (circles)
are compared with those obtained from Eq. (\ref{e2}) (line).
The agreement is again very good up to very high-spins.
Note that the accuracy of the $1/N$ expansion results improves with increasing
$N$, hence in actual applications with $N\sim 30$, one would expect an even
better agreement.
The test case discussed here indicates that the extended formalism can be
applied with confidence in the spin region $L=N$-$3N$ which covers the
presently available data range for superdeformed bands.

\section{APPLICATIONS TO SUPERDEFORMED BANDS}

In this section, we apply the $1/N$ expansion formulas first in a
systematic study of dynamic moment of inertia and $B(E2)$ values, and then
to describe the experimental data on superdeformed bands.
Since $\kappa$ is a scale parameter for energies, we need to study
the effect of the three quadrupole parameters $q_{22}$, $q_{24}$, $q_{44}$.
Fig.~\ref{fig3} shows the effect of variations in each $q_{jl}$ on
dynamic moment of inertia while the other two are held constant at the
SU(3) values. Here $q$ denotes the scaling parameter from the SU(3) values.
Thus $q=1$, corresponds to the SU(3) limit which exhibits the rigid-rotor
behaviour. To describe the variations in ${\cal J}^{(2)}$, one needs to break
the SU(3) limit. From Fig.~\ref{fig3} it is seen that ${\cal J}^{(2)}$
is most sensitive to $q_{24}$ (note the different scales in the three figures).
The other (diagonal) parameters have smaller and opposite effect on ${\cal
J}^{(2)}$. Since the amount of data does not justify use of too many
parameters, we prefer to scale all three with the same parameter $q$. The
result of this simultaneous scaling is shown in Fig.~\ref{fig4}-a which is
essentially the same as the one for $q_{24}$ in Fig.~\ref{fig3}. An interesting
feature of these results is that the quadrupole Hamiltonian has the scope to
describe both the increases and decreases in ${\cal J}^{(2)}$. For $q<1$,
the $s-d$ coupling is relatively stronger than the $d-g$ coupling which results
in loss of monopole pairing with increasing spin, and hence increase in ${\cal
J}^{(2)}$. The opposite happens for $q>1$.
Fig.~\ref{fig4}-b shows the effect the simultaneous variations in the
quadrupole parameters on the $B(E2)$ values. The curving down of lines is due
to boson cutoff which is most effective for smaller values of $q$.

In the light of the above systematic studies, we have carried out fits to the
available data on superdeformed bands in the Hg-Pb and Gd-Dy regions.
The boson number is determined from microscopics, and $\kappa$ and $q$ are
fitted to the data. The  parameter values are given in the figure
captions and the data are taken from the compilation in Ref. \cite{fir94}.
Figs.~\ref{fig5} and \ref{fig6} compare the experimental dynamic moment of
inertia (circles) with the calculated ones (lines) in Hg and Pb isotopes,
respectively. In all cases ${\cal J}^{(2)}$ exhibits a smooth increase which
is well reproduced by the calculations.
The situation in the Gd-Dy region is not as favorable for our simple
collective model as the other region, because
 there are definite signs indicating the importance of the
single particle degree of freedom.
For example, in $^{144-146}$Gd there are
sudden jumps in ${\cal J}^{(2)}$ which are probably
due to particle alignment effects
\cite{lun94}. In $^{148-150}$Gd, ${\cal J}^{(2)}$ behaves reasonably smoothly
so we have attempted to describe them (see Fig.~\ref{fig7}). The average
behaviour of ${\cal J}^{(2)}$ in $^{148}$Gd is reproduced but the model fails
in the case of $^{150}$Gd, underscoring the importance of single particle
degree of freedom. For a better description of the data, one needs to
incorporate particle alignment effects in the present formalism by including
two-quasiparticle states in the basis \cite{iac91}.
The dynamic moment of
inertia of superdeformed bands in Dy isotopes exhibit an entirely different
behaviour (Fig.~\ref{fig8}). They are very close to the rigid-rotor values, and
hence the SU(3) limit as reflected in the values of $q\sim 1$.

The $B(E2)$ values provide a complimentary observable to ${\cal J}^{(2)}$ which
could be used as a further test of the model.
In Fig.~\ref{fig9}, the available $B(E2)$ data in $^{192-194}$Hg (circles) are
compared with the calculations. A reasonable description is obtained using
boson effective charges $e=0.12-0.14~eb$ which are typical values used in the
normal IBM calculations. Since the $B(E2)$ values are sensitive to the boson
number (they vary as $N^2$), this provides a consistency check on the
microscopically derived boson numbers. A further $N$ dependence is provided by
the boson cutoff term in Eq. (\ref{e2}) which causes a drop in the calculated
$B(E2)$ values at high-spins. Least-square fits to the data indeed indicate a
drop in the $B(E2)$ values towards the high-spin end. However, the error bars
are too large to reach an unambiguous conclusion whether this effect is genuine
or not.

\section{CONCLUSIONS}
In this paper, we have reviewed a microscopic basis and a practical
formulation of the IBM for application to superdeformed nuclei. The
availability of analytical formulas owing to the 1/$N$ expansion technique
 means fast and efficient analysis of data.
As first examples, we have considered the superdeformed bands in the Hg-Pb and
Gd-Dy regions. A good description of data is obtained in the Hg-Pb region
confirming the simple quadrupole nature of these superdeformed bands.
In the Gd-Dy region, the dynamic moment of inertia exhibits large variations
which can not be accommodated in a simple collective model. Such variations are
due to the single particle degree of freedom and require extension of the model
for a better description of the data.
Finally, the formalism can be used in investigating some other interesting
features of superdeformed nuclei such as identical bands and $C_4$ symmetry
which will be pursued in future work.

\section{ACKNOWLEDGEMENTS}

This work is supported in parts by an exchange grant from the Australian
Academy of Science/Japan Society for Promotion of Science and by the
Australian Research Council, and in parts by Grant-in-Aid for Scientific
Research on Priority Areas (No. 05243102) from the Ministry of Education,
Science and Culture.

\begin{figure}
\caption{Occupation probability of each spherical basis  in the Nilsson
potential: (a) protons and (b) neutrons of $^{194}$Hg  and (c) protons and (d)
neutrons of $^{152}$Dy.  Two values of  the deformation parameter are
considered, $\delta=-0.13$ (dashed line)  and 0.40 (solid line)  for
$^{194}$Hg, and $\delta=0.25$ (dashed line) and 0.50 (solid line)  for
$^{152}$Dy.  The spherical magic numbers are indicated.}
\label{fig1}
\end{figure}

\begin{figure}
\caption{Comparison of the 1/$N$ expansion results (lines) with the exact
numerical ones (circles) for $N=10$ bosons.
Fig. 2-a shows the dynamic moments of inertia obtained from the third order
calculation with VAP (solid line), third order with VBP (dashed line), and
second order with VAP (dotted line). The parameters of $H$ are $\kappa = 20$
keV and $q = 0.7$. Fig. 2-b compares the $E2$ matrix elements.}
\label{fig2}
\end{figure}

\begin{figure}
\caption{Systematic behaviour of dynamic moment of inertia
${\cal J}^{(2)}$ for various values
of the quadrupole parameters $q_{22}$, $q_{24}$ and $q_{44}$. In each figure,
two of these three parameters are fixed at the SU(3) value while the
 other takes 0.6, 0.8,
1.0, 1.2, 1.4 times the SU(3) value. $N=30$ and $\kappa$ is chosen such that
the moment of inertia are all normalized to 100$\hbar^{2}$MeV$^{-1}$
 at $\omega=0$.}
\label{fig3}
\end{figure}

\begin{figure}
\caption{Effect of the simultaneous scaling of the quadrupole parameters
(a) on dynamic moment of inertia, and (b) on $B(E2)$ values.
The dynamic moment of inertia curves are normalized to 100
$\hbar^{2}$MeV$^{-1}$ at $\omega=0$ and
the $B(E2)$ values are normalized to $B(E2; 2\to 0)$.}
\label{fig4}
\end{figure}

\begin{figure}
\caption{Comparison of the experimental dynamic moment of inertia in
$^{190-194}$Hg (circles) with the super IBM calculations (solid lines).
$N=29, 30, 31$, $\kappa= 35, 34, 33$ keV, $q=0.68, 0.72,
0.72$ are taken for $^{190-194}$Hg, respectively. The data are from \protect
\cite{fir94}.}
\label{fig5}
\end{figure}

\begin{figure}
\caption{Same as Fig. 5 but for $^{192-196}$Pb.
$N=30, 31, 32$, $\kappa= 33, 33, 34$ keV,
$q=0.66, 0.65, 0.67$ are taken for $^{192-196}$Pb, respectively.}
\label{fig6}
\end{figure}

\begin{figure}
\caption{Same as Fig. 5 but for $^{148-150}$Gd.
$N=29, 30$, $\kappa= 41, 27$ keV, $q=1.35, 1.73$
are taken for $^{148-150}$Gd, respectively.}
\label{fig7}
\end{figure}

\begin{figure}
\caption{Same as Fig. 5 but for $^{152-154}$Dy.
$N=31, 32$, $\kappa= 42, 43$ keV, $q=1.07, 1.02$
are taken for $^{152-154}$Dy, respectively.}
\label{fig8}
\end{figure}

\begin{figure}
\caption{Comparison of the experimental $B(E2)$ values in
$^{192-194}$Hg (circles) with the super IBM calculations (solid lines).
The boson effective charges are $e=0.140$ and 0.124~$eb$ for $^{192-194}$Hg,
respectively.
The data are from \protect \cite{hug94,wil94}.}
\label{fig9}
\end{figure}

\mediumtext
\begin{table}
\caption{Probability (\%) of each angular momentum component
 in the ${\it \Lambda}$-pair.
 The ${\it \Lambda}$-pair is obtained from the Nilsson + particle number
 conserving BCS wave function.
Two cases of $\delta=-0.13$ and $0.40$ are shown for $^{194}$Hg.
The probability of each boson in the intrinsic boson
of the IBM in the SU(3) limit is also listed for comparison.}
\label{spindec}
\begin{center}
\begin{tabular}{crrcrrcrr}
 pair/boson & \multicolumn{2}{c}{$\delta=-0.13$} &
 & \multicolumn{2}{c}{$\delta= 0.40$} &
 & \multicolumn{2}{c}{IBM-SU(3)} \\ \cline{2-3} \cline{5-6} \cline{8-9}
 & \multicolumn{1}{c}{neutron} & \multicolumn{1}{c}{proton} &
 & \multicolumn{1}{c}{neutron} & \multicolumn{1}{c}{proton} &
 & \multicolumn{1}{c}{sd} & \multicolumn{1}{c}{sdg} \\ \hline
$S/s$ & 82 & 93 & & 25 & 32 & & 33  & 20 \\
$D/d$ & 18 &  7 & & 52 & 51 & & 67  & 57 \\
$G/g$ &  0 &  0 & & 18 & 14 & & $-$ & 23 \\
\end{tabular}
\end{center}
\label{table1}
\end{table}

\end{document}